\documentclass[trackchanges, twocolumn]{aastex7}

\shorttitle{RRAT~J1541+4703}
\shortauthors{Xin Xu \& Qijun Zhi et al.}
\graphicspath{{./}{figures/}}
\usepackage{hhline}

\begin{document}


\title{RRAT~J1541+4703: A Rotating Radio Transient Exhibiting Normal Pulsar States}


\correspondingauthor{Qijun Zhi}\email{qjzhi@gznu.edu.cn}
\correspondingauthor{Shijun Dang}\email{dangsj@gznu.edu.cn}

\author[orcid=0009-0006-3224-4319]{Xin Xu}
\affiliation{School of Mathematical Science, Guizhou Normal University, Guiyang 550001, People’s Republic of China}
\affiliation{Guizhou Provincial Key Laboratory of Radio Astronomy and Data Processing, Guizhou Normal University, Guiyang 550001, People’s Republic of China}
\email{xuxin_nl@163.com}

\author[orcid=0000-0001-9389-5197]{Qijun Zhi}
\affiliation{School of Science, Guizhou University, Guiyang 550001, People’s Republic of China}
\affiliation{School of Physics and Electronic Science, Guizhou Normal University, Guiyang, 550001, People’s Republic of China}
\affiliation{Guizhou Provincial Key Laboratory of Radio Astronomy and Data Processing, Guizhou Normal University, Guiyang 550001, People’s Republic of China}
\email{qjzhi@gznu.edu.cn}

\author{Jie Tian}
\affiliation{School of Mathematical Science, Guizhou Normal University, Guiyang 550001, People’s Republic of China}
\affiliation{Guizhou Provincial Key Laboratory of Radio Astronomy and Data Processing, Guizhou Normal University, Guiyang 550001, People’s Republic of China}
\email{jietian2024@163.com}

\author[orcid=0000-0002-9815-5873]{Jiguang Lu}
\affiliation{Guizhou Radio Astronomical Observatory, Guiyang 550025, People’s Republic of China}
\affiliation{National Astronomical Observatories, Chinese Academy of Sciences, Beijing 100012, People’s Republic of China}
\email{lujig@nao.cas.cn}

\author[orcid=0000-0002-5387-7952]{Peng Jiang}
\affiliation{Guizhou Radio Astronomical Observatory, Guiyang 550025, People’s Republic of China}
\affiliation{National Astronomical Observatories, Chinese Academy of Sciences, Beijing 100012, People’s Republic of China}
\email{pjiang@nao.cas.cn}

\author[orcid=0000-0002-2060-5539]{Shijun Dang}
\affiliation{Guizhou Provincial Key Laboratory of Radio Astronomy and Data Processing, Guizhou Normal University, Guiyang 550001, People’s Republic of China}
\affiliation{School of Physics and Electronic Science, Guizhou Normal University, Guiyang, 550001, People’s Republic of China}
\email{dangsj@gznu.edu.cn}

\author[orcid=0000-0002-9042-3044]{Renxin Xu}
\affiliation{School of Physics and State Key Laboratory of Nuclear Physics and Technology, Peking University, Beijing 100871, China}
\email{r.x.xu@pku.edu.cn}

\author[orcid=0000-0002-1052-1120]{Juntao Bai} 
\affiliation{Institute for Gravitational Wave Astronomy, Henan Academy of Sciences, Zhengzhou 450046, Henan, People’s Republic of China}
\email{jtbai@hnas.ac.cn}

\author{Ke Yang}
\affiliation{Guizhou Provincial Key Laboratory of Radio Astronomy and Data Processing, Guizhou Normal University, Guiyang 550001, People’s Republic of China}
\affiliation{School of Physics and Electronic Science, Guizhou Normal University, Guiyang, 550001, People’s Republic of China}
\email{yangk1632617@163.com}

\collaboration{all}{}


\begin{abstract}
Rotating Radio Transients (RRATs) are a class of pulsar-like objects characterized by intermittent radio emissions. Among them, RRATs that exhibit both RRAT and normal pulsar (NP) states may represent a key evolutionary stage from nulling pulsars to RRATs. 
We performed a detailed analysis of RRAT J1574+4703 using the Five-hundred-meter Aperture Spherical Radio Telescope (FAST) at a frequency of 1250 MHz. Our findings indicate that this RRAT spends approximately 98\% of its time in the RRAT state, with the remainder spent in an NP state exhibiting nulling behavior. Additionally, we observed distinct integral pulse profiles and polarization properties between the two states, suggesting that they originate from different emission heights and magnetospheric structures. 
Furthermore, it was observed that the NP states of this RRAT exhibit mode switching, with $\sim 44\%$ of the time spent in the normal mode and $\sim 39\%$ in the abnormal mode. Notably, abnormal modes are predominantly detected at the onset and termination of the NP states. This discrepancy between the modes indicates potential instability in the magnetospheric processes that govern the NP states.

\end{abstract}





\section{Introduction} \label{sec_intro}

Rotating Radio Transients (RRATs) are a class of celestial objects known for their intermittent electromagnetic emission. Their radio emissions are brief, typically lasting only a few milliseconds, followed by quiet periods that can last from minutes to hours  \citep{2011MNRAS.415.3065Keane}. 
This character makes it challenging to detect their periodicity through traditional Fourier domain searches; that is why RRATs are more commonly identified through single-pulse searches. The first RRAT was discovered using this method  \citep{2006Natur.439..817McLaughlin}. 
Despite over 150 RRATs having been detected to date \footnote{https://rratalog.github.io/rratalog/}, their number remains considerably lower than the known pulsar population \citep{2005AJ....129.1993Mancheste}.
Some studies suggest that RRATs may represent a subclass of radio pulsars, potentially linked to other pulsar populations. For instance, the thermal radiation of RRATs resembles that of cooling neutron stars \citep{2006ApJ...639L..71Reynolds} , and their pulse energies follow a log-normal distribution \citep{2017ApJ...840....5CuiBY}, similar to normal pulsars. Additionally, RRATs exhibit pulsar-like post-glitch timing properties after glitches \citep{2018MNRAS.477.4090Bhattacharyya}. Therefore, in-depth studies of RRATs with unique phenomena are essential to explore further connections between RRATs and normal pulsars (NP).

Notably, some studies have detected continuous pulse sequences in certain RRATs, such as J1913+1330 and J1538+2345 \citep{2019SCPMA..6259503LuJG, 2024MNRAS.527.4129ZhongWQ}. Some RRATs display both continuous and sporadic pulse sequences, alternating between pulsar and RRAT states, as seen in J0828-3417, J0941-39, and PSR J1107-5907 \citep{2010MNRAS.402..855Burke, 2012ApJ...759L...3Esamdin, 2023ApJ...959...56SunSN}. This supports the hypothesis that RRATs could be extremely nulling pulsars. 
The null fraction (NF) of pulsars is positively correlated with their characteristic age \citep{2007MNRAS.377.1383WangN}, as older pulsars can no longer sustain pair-production processes in their magnetospheres, leading to nulling behavior \citep{1975ApJ...196...51Ruderman, 1997ApJ...478..313ZhangB}. 
As a pulsar ages, the NF of pulses may rise, potentially reaching the extreme nulling state seen in RRATs \citep{2010MNRAS.402..855Burke}. 
RRATs that exhibit both pulsar and RRAT states could represent an important evolutionary transition from nulling pulsars to RRATs. Further studies of these objects will improve our understanding of RRAT emission mechanisms and their connection to pulsars.

RRAT~J1541+4703 was discovered by \citet{2023MNRAS.524.5132Dong} using the Canadian Hydrogen Intensity Mapping Experiment (CHIME). It has a period of $\rm P = 0.2777 s$ and a period derivative of $2.098\times10^{-16} \rm s\,s^{-1}$. The characteristic age of this RRAT is estimated to be $\rm 2.1 \times 10^{7} Yr$, which is older than that of typical pulsars. Its surface magnetic field strength is $\rm  B=2.44 \times 10^{11}G$, indicating a stronger magnetic field compared to normal pulsars. 
Using the CHIME/Pulsar instrument, \citet{2023MNRAS.524.5132Dong} conducted about 79 hours of observations of this RRAT, detecting 639 bursts, with a pulse rate of approximately $\rm \sim 8 hr^{-1}$. 
The observations revealed that this RRAT exhibits subpulse drifting, extremely narrow pulses, and often displays double peaks. 
This suggests that RRAT~J1541+4703 exhibits unique emission phenomena. However, detailed studies of this RRAT are currently lacking. 
The Five-hundred-meter Aperture Spherical Radio Telescope (FAST), with its ultra-high sensitivity \citep{2011IJMPD..20..989NanRD}, can observe finer details. 
It has already detected numerous phenomena that remain unclear to other telescopes \citep[e.g. ][]{2023MNRAS.520.1332ZhiQJ, 2024MNRAS.527.3761XuX, 2024ApJ...968..119XuX}, as well as discovering many weak pulsars \citep{2024ApJ...960...79ZhiQJ, 2025ApJ...982..117XuX, 2025RAA....25a4001HanJL}.

In this paper, we present results from observations of RRAT~J1541+4703 using FAST at a center frequency of 1250 MHz. The findings indicate that this RRAT simultaneously exhibits both pulsar and RRAT states, with mode changing detected in its NP state. 
To provide a comprehensive account of this work, the remainder of this paper is organized as follow. Section \ref{sec_obs} presents the observations of RRAT~J1541+4703 with FAST, along with the data reduction methods. The results and analysis are provided in Section \ref{sec_res}, while discussions and conclusions are presented in Section \ref{sec_dis_con}.

\begin{table*}
\centering
\caption{Details of the observations and separation results for RRAT~J1541+4703.  \label{table_obspara}}
{\begin{tabular}{cccccccccccccc}
 \hline
 \hline
 Epoch & Date & Period epoch & Observing length & $\rm Zap_{chan}$ & \multicolumn{6}{c}{Number of Pulse} & Pulse Rate of \\
 \cline{6-11}
  & (y-m-d) & (MJD) & (min) & (\%) & Total & \multicolumn{2}{c}{RRAT} && \multicolumn{2}{c}{Pulsar} & RRAT state \\
 \cline{7-8}
 \cline{10-11}
  & & & & & & Burst & Null && Burst & Null & (hr$^{-1}$)\\
 \hline
 1 & 2024-09-05 & 60558 & 21 & 13.0 & 4191 & 10 & 4137 && 36 & 8 & 31  \\
 2 & 2024-10-16 & 60599 & 55 & 13.0 & 11537 & 261 & 11134 && 114 & 28  & 297  \\
 3 & 2024-11-07 & 60621 & 60 & 9.5 & 12617 & 394 & 11948 && 240 & 35  & 414  \\
 4 & 2024-12-13 & 60657 & 130 & 13.4 & 27742 & 114 & 27424 && 132 & 72  & 54  \\
 5 & 2025-06-04 & 60830 & 60 & 12.2 & 12616 & 388 & 12145 && 70 & 13  & 401  \\
 6 & 2025-06-27 & 60853 & 60 & 15.6 & 12570 & 325 & 12022 && 184 & 39  & 341  \\
 7 & 2025-07-09 & 60865 & 60 & 7.3 & 12617 & 549 & 11528 && 473 & 67  & 589  \\     
 \hline
 \hline	
\end{tabular}}
\end{table*}


\section{Observations and Data Reduction} \label{sec_obs}

We observed RRAT~J1541+4703 using the central beam of FAST's 19-beam L-band receiver in "swiftcalibration" mode. Seven observations were conducted, lasting between 21 and 130 minutes. The observation times and durations for each epoch are listed in Table~\ref{table_obspara}. Observations were made at a center frequency of 1.25 GHz, with a 500 MHz bandwidth spread across 4096 frequency channels. The 50 MHz at the edges was flagged and removed, leaving an effective bandwidth of 1.05 to 1.45 GHz  \citep{2019SCPMA..6259502JiangP}. Data were recorded in search mode using the PSRFITS format \citep{2004PASA...21..302Hotan}. The time resolution of 49.152~$\rm \mu$s. 
Before each observation, the telescope was pointed to a non-source location for polarization calibration, where a one-minute noise diode signal was injected at $\rm 45^\circ$. The telescope was then directed to the this RRAT's location. The number of pulses recorded per epoch is listed in the fifth column of Table~\ref{table_obspara}.

The data were processed into single pulse archives using the DSPSR software package \citep{2011PASA...28....1Van_Straten}, with each pulse period divided into 512 bins based on the timing ephemeris from the ATNF pulsar catalog \citep[PSRCAT]{2005AJ....129.1993Manchester}, and inter-channel dispersion delays were eliminated. 
We then employed the PAZ program from the PSRCHIVE software package \citep{2004PASA...21..302Hotan, 2012AR&T....9..237Van_Straten} to remove radio frequency interference (RFI) from the dispersed single pulse data. The procedure implemented an automatic channel zapping algorithm based on a tolerance threshold for deviations between the bandpass and its median smoothed version. A median smoothing window of 21 channels was applied, and any frequency channel with a total flux exceeding 4 standard deviations from the median had its weight set to zero. 
Additionally, the interactive PAZI tool in PSRCHIVE was used to manually flag and remove obvious narrowband RFI, minimizing its impact on the pulse signal. The percentage of frequency channels removed in each observation is listed in the fifth column of Table~\ref{table_obspara}.
It is worth noting that this method primarily identifies RFI based on deviations from the bandpass in the frequency domain. For the detection of weak and sporadic pulses, a systematic time–frequency domain analysis to identify and remove interference could further improve data quality in future studies \citep{2021ApJ...918...57WenZG}. 
For polarization calibration, noise diode observations were performed before each observation. During this process, the telescope pointed 10 arcminutes away from the target source and injected a 100\% linearly polarized diode signal  \citep{2020RAA....20...64JiangP}. Specific parameters, such as period and duration, can be found on the FAST website~\footnote{https://fast.bao.ac.cn/}. 
Polarization calibration for each observation was then completed using the PAC program in PSRCHIVE. Subsequently, the RMFIT program in PSRCHIVE was used to search for the Faraday rotation measure (RM) of RRAT~J1541+4703 within an RM range of $\pm 1000 \,\rm rad\,m^{-2}$. The result indicates that the RM value for this RRAT is $9.5 \pm 3.8 \,\rm rad\,m^{-2}$. 
Finally, the RM value was applied to eliminate for ionospheric and interstellar medium effects during pulse propagation, thereby correcting the pulse polarization profile.

Since our observations did not undergo flux calibration, the peak flux density of each burst was estimated using the radiometer equation to convert the data to the Jansky scale \citep{2017ApJ...840....5CuiBY}:
\begin{equation}
    \rm S_{peak} = \frac{A_{peak} T_{sys}}{\sigma G \sqrt{t_{obs} n_p \Delta \nu}}
    \label{Eq_flux}
\end{equation}
where $\rm A_{peak}$ is the peak within the pulse window, and $\rm \sigma$ is the root mean square value of the phase amplitude outside the pulse window. $\rm T_{sys} = 20 K$, $\rm G = 16 K Jy^{-1}$, $\rm n_p = 2$, $\rm \Delta \nu = 400~MHz$ and $\rm t_{obs}$ are the system-noise temperature, the gain of the telescope, dual-polarization data, the observing bandwidth and observation sample time, respectively \citep{2019SCPMA..6259502JiangP}.


\section{Results and Analysis} \label{sec_res}

\subsection{Burst and null pulses separation} \label{subsec_Det} 

A null pulse in RRATs occurs usually when their emission ceases entirely or becomes too weak to be detected \citep{2006ApJ...645L.149Weltevrede, 2007MNRAS.374.1103ZhangB}. The high sensitivity of FAST greatly improves the detection of such weak signals. To analyze the emission characteristics of RRAT~J1541+4703 during burst states, we labeled all single pulses, distinguishing between null and burst pulses. While some studies use the peak signal-to-noise ratio to identify burst pulses \citep[e.g. ][]{2012ApJ...759L...3Esamdin, 2023ApJ...959...56SunSN, 2025ApJ...988...11DangSJ}, J1541+4703 exhibits narrow and weak pulses \citep{2023MNRAS.524.5132Dong}, which require a more sensitive approach. 
Therefore, we employed the \citet{2023NatAs...7.1235ChenX}’s method. The $\rm W_{10}$ of the integrated pulse profile defined the detection window. 
Within this window, pulses with flux exceeding specified threshold combinations ($n_1$, $n_2$, $n_3$)$\sigma_{\rm bin}$ for three consecutive, two consecutive, or a single phase bin are classified as burst pulses. Here, $n_1$, $n_2$, and $n_3$ represent the thresholds for three consecutive, two consecutive, and single phase bins, respectively, and $\sigma_{\rm bin}$ is the standard deviation of the amplitude in the off-pulse region, defined to match the width of $\rm W_{10}$. Pulses that do not meet any of these criteria are classified as null pulses.

\begin{figure} 
 \centering
 \includegraphics[width=0.49\textwidth]{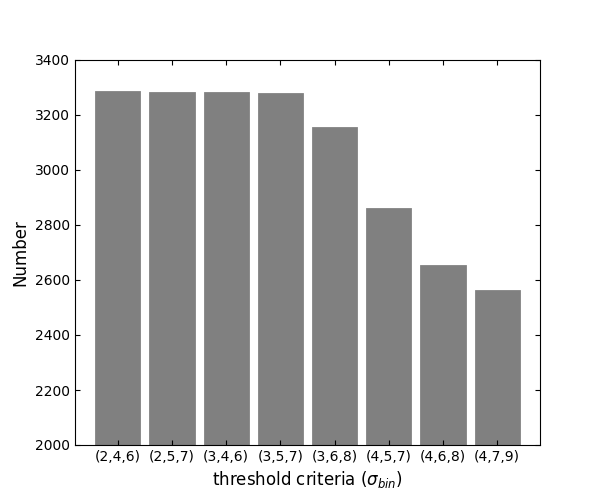}
 \caption{
 Number of burst pulses detected under different threshold combinations. 
 \label{fig_threshold_criteria}} 
\end{figure}

Since burst pulse detection depends on the chosen threshold criteria, we applied different threshold combinations and examined the corresponding frequency–phase plots to verify that the detected signals exhibit genuine broadband emission. This allowed us to determine the number of burst pulses detected under each threshold combination, as shown in Figure~\ref{fig_threshold_criteria}. 
The results indicate that as thresholds decrease, the number of detected pulses increases and eventually stabilizes. However, excessively low thresholds result in misclassification of null pulses as burst pulses, while overly high thresholds cause some weak pulses to be missed. Based on this trade-off, we adopt the threshold combination $(3,5,7)\sigma_{\rm bin}$ in this study.
We note that a small number of weak single pulses maybe misclassified. Extremely narrow pulses may fail toreach the single-bin threshold of $7\sigma_{\rm bin}$, while very broadand weak pulses may not satisfy the criterion of threeconsecutive phase bins above threshold of $3\sigma_{\rm bin}$. 
To quantitatively assess the reliability of our detection method, we injected $10^5$ simulated pulses with various widths and peak fluxes into off-pulse noise and processed them through the above method. The overall recovery rate reaches 97.4\%, demonstrating that our method reliably captures the vast majority of burst pulses.
Matched-filtering techniques, which are more sensitive to weak and narrow pulses, may further validate the burst pulses of this RRAT in future studies. Additionally, observations with higher sensitivity could improve the detection rate, though this does not affect the main conclusions of this work.

Using this method, we classified all single pulses. The number of null and burst pulses for each observation is listed in columns {6 to 9 of Table~\ref{table_obspara}, with separate counts for RRAT and NP states. These two states are discussed in Section~\ref{subsec_twostate}. Epochs 1 and 4 experienced significant interference, which obscured weaker bursts, making them undetectable. Consequently, fewer burst pulses were detected during these epochs compared to others in the same timeframe.

\begin{figure*} 
 \centering
 \includegraphics[width=0.99\textwidth]{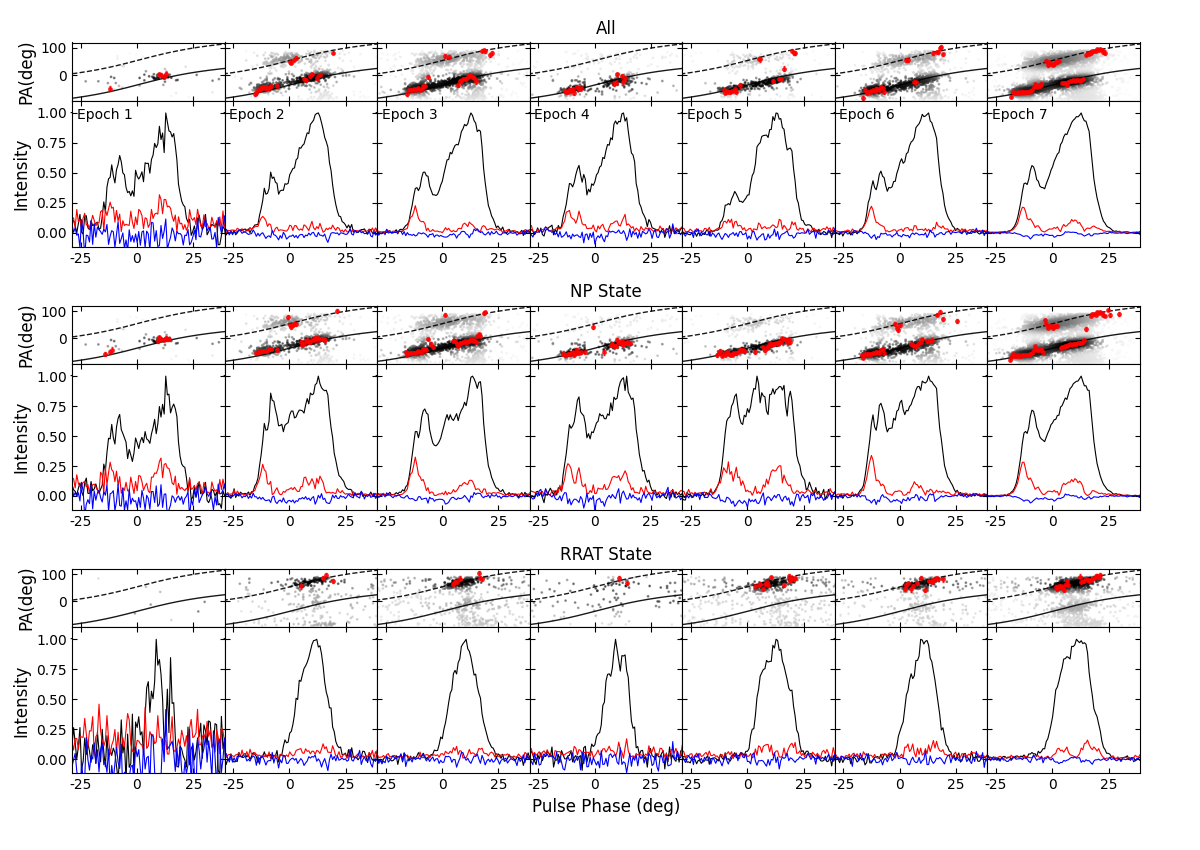}
 \caption{
 The polarized integrated pulse profile of RRAT~J1541+4703 from seven epochs, including all burst pulses, NP states, and RRAT states. In the top panel of each row, red and gray dots denote the PA of the integrated pulse profile and that of a single pulse, respectively. The solid and dashed black lines indicating RVM fitting curves separated by $90^{\circ}$. The corresponding bottom panel shows total intensity (black line), linear polarization (red line), and circular polarization (blue line). All pulse profiles have been normalized to peak intensity. 
 \label{fig_mean_PA}} 
\end{figure*}


\subsection{Polarization Profiles} \label{subsec_pol}

Due to the sparse bursting nature of RRATs, the number of null pulses significantly outweighs the burst pulses. As a result, the signal-to-noise ratio of the integrated pulse profile, obtained by summing all single pulses, is low. To address this, we accumulated only the burst pulses to derive the integrated pulse profile. The first row of Figure~\ref{fig_mean_PA} shows the polarization integrated pulse profile and polarization angle (PA) for RRAT~J1541+4703 at different epochs.

All observational results were combined to obtain a high signal-to-noise ratio integrated pulse profile of RRAT~J1541+4703, shown by the black curve in Figure~2. The profile was shifted so that the phase of maximum RVM slope corresponds to $0^{\circ}$.
Using the method described by \citet{1994A&AS..107..527Kramer}, and assuming each component follows a Gaussian shape, we fit the profile with two or three Gaussian functions:
\begin{equation}
  \rm Y = \sum_{i=1}^{n} I_i exp \left[-4 ln2 \left(\frac{x-x_i}{w_i} \right)^2 \right]
  \label{Eq_guass}
\end{equation}
where $\rm I_i$ is the amplitude of the $\rm i$th Gaussian component, $\rm x_i$ is the peak position, and $\rm w_i$ is the full width at half maximum, n is number of component. 
The three-component fit is shown in Figure~\ref{fig_guassfit}. Its residuals exhibit noise-like behavior comparable to the off-pulse region, whereas systematic structures remain when only two Gaussian components are used. The fitting parameters are listed in Table~\ref{table_guass}. In addition, the three-component model yields a smaller $\chi^2$ value than the two-component model. These results demonstrate that the integrated profile of this RRAT is well fitted by the three Gaussian components, while small-scale features are likely attributable to noise. 
We further evaluated both models using the Akaike Information Criterion (AIC) and the Bayesian Information Criterion (BIC). The three-component model yields lower values (AIC = 296.28, BIC = 322.34) than the two-component model (AIC = 498.66, BIC = 516.89), again favoring the three-component interpretation.
The results indicate that the second component is not centered within the profile but is shifted toward the tail. It partially overlaps with the third component, making it challenging to distinguish from the integrated profile alone. This asymmetry is observed in the radiation of many pulsars \citep{1983ApJ...274..333Rankin}. 
The amplitude of the first component is approximately half the peak height of the integrated pulse profile. We measured the pulse widths at 10\% and 50\% of the peak, yielding $\rm W_{10} = 37.9^{\circ} \pm 4.6^{\circ}$ and $\rm W_{50} = 21.2^{\circ} \pm 0.5^{\circ}$, respectively. Linear polarization is more prominent in the leading part of the profile than in the trailing part, with the overall polarization fraction averaging 12.9\% ± 2.1\%.

The average PA and single pulse PA of this RRAT are shown in the top panel of the first row of Figure~\ref{fig_mean_PA}, represented by red error bars and gray dots, respectively.
The results reveal that, except for Epochs 1 and 4, the PA in all other epochs shows two $\sim 90^{\circ}$ jumps, occurring around pulse phases $\sim 0^{\circ}$ and $\sim 20^{\circ}$, suggesting a discontinuous orthogonal polarization mode (OPM). We attribute the lack of detectable PA values at these phases in Epochs 1 and 4 to the low signal-to-noise ratio.

To determine the geometric parameters of this RRAT, we performed the rotating vector model  \citep[RVM, ][]{1969ApL.....3..225Radhakrishnan} fiting on the integrated pulse polarization profiles obtained from all observations:
\begin{equation}
    \rm \tan(\psi-\psi_0) = \frac{\sin \alpha \sin(\phi-\phi_0)}{\sin (\zeta) \cos \alpha - \cos (\zeta) \sin \alpha \cos(\phi-\phi_0)}
    \label{Eq_RVM}
\end{equation}
where $\zeta = \alpha + \beta$, which is the angle between the line of sight and the rotation axis, with $\alpha$ being the magnetic inclination angle and $\beta$ the impact parameter. $\psi$ is the PA at a pulse phase $\phi$, and $\psi_0$ and $\phi_0$ are the phase offsets for PA and pulse rotational phase, respectively. 
During the jump phase, the PA values were shifted by $90^{\circ}$. This method is widely used to fit RVM models for pulsars with frequent PA jumps \citep[e.g. ][]{2019MNRAS.490.4565Johnston, 2023ApJ...949..115YuanM, 2025ApJ...989..127XuX}.

We used the Python package EMCEE \citep{2013PASP..125..306Foreman-Mackey} to fit a Markov chain Monte Carlo (MCMC) model \citep{2019MNRAS.490.4565Johnston} to the PA variations of this pulsar and determine these parameters, as described by \citet{2025ApJ...989..127XuX}. 
In the MCMC fitting, all parameters are assigned a noninformative uniform prior distributions. Their corresponding posterior distributions are shown in Figure \ref{fig_mcmc}.
Each parameter value is determined from the median of the posterior distribution, with uncertainties estimated from the 16th and 84th percentiles. 
The best-fitting values are $\rm \alpha = 87.3^{+38.8}_{-32.4}$, $\rm \beta = 27.3^{+3.0}_{-8.0}$, $\rm \phi_0 = 0.0^{+2.7}_{-3.6}$ and $\rm \psi_0-34.2^{+4.7}_{-6.4}$.
The results suggest that this RRAT has a larger inclination angle.

Under the RVM framework, when $\alpha$ nears $90^\circ$, the solution for $\beta$ tends to degenerate between positive and negative values. We performed MCMC fitting with a noninformative uniform prior for $\beta$ over the symmetric interval [$-90^\circ$, $90^\circ$]. The results revealed a single distribution centered around $27^\circ$, indicating that the $\beta$ parameter for this RRAT shows a clear bias toward the positive solution.
The best RVM curve is shown as the solid black curve in the top panel of each row in Figure~\ref{fig_mean_PA}, with the dashed black curve representing the result of shifting the solid curve upward by $90^{\circ}$.

Our RVM fitting suggests that J1541+4703 may be an orthorotating pulsar with a $\alpha$ close to $90^{\circ}$ and $\beta$ around $27^\circ$. Under these conditions, the line of sight could intersect two radiation beams from opposite magnetic poles, potentially leading to the observation of an interpulse at a phase angle approximately $180^\circ$ separated from the main pulse. 
We examined the entire pulse period and found no evidence of an interpulse, which may be too weak to detect. Assuming the width of the potential interpulse is identical to that of the main pulse, we estimate the $3\sigma_0$ fluence upper limit for the potential interpulse to be around 0.03\% of the main pulse, where $\sigma_0$ is the standard deviation of the amplitude in interpulse region. The existence of the interpulse in this RRAT requires further high-sensitivity observations for confirmation.

\begin{figure} 
 \centering
 \includegraphics[width=0.49\textwidth]{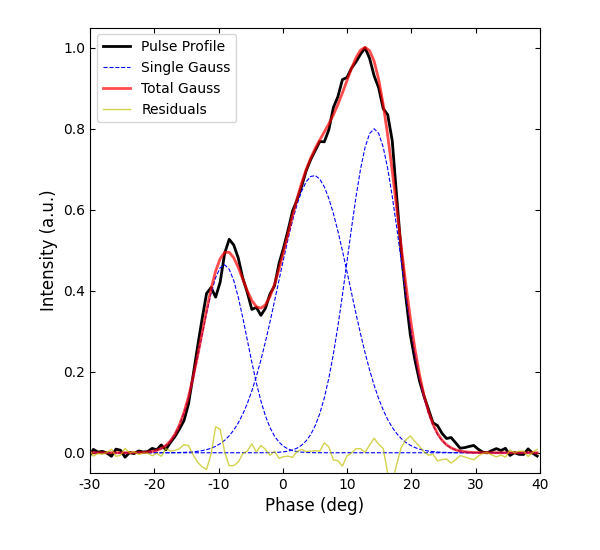}
 \caption{
 Results of the three Gaussian fits for the integrated pulse profile. The black solid line represents the integrated pulse profile, the red solid line the fitted curve, and the blue dashed line the single Gaussian component. The residuals of the fit are represented by the yellow curves.
 \label{fig_guassfit}} 
\end{figure}

\begin{figure} 
 \centering
 \includegraphics[width=0.48\textwidth]{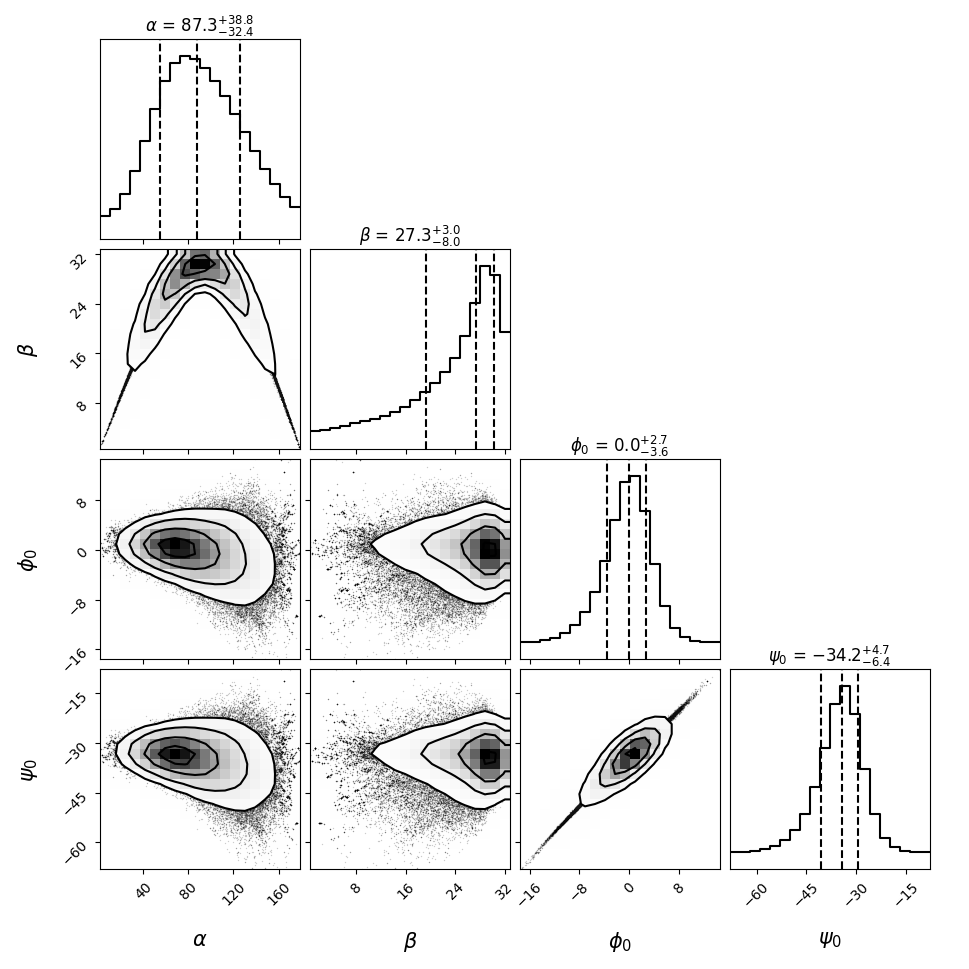}
 \caption{
 An example of the posterior distributions for four parameters in RVM model fitting. The contours represent $1 \sigma$, $2 \sigma$, and $3 \sigma$ confidence levels. The vertical dashed lines in the top panel of each column indicate the 16th, median, and 84th percentiles of the distribution. The fitted values and their errors for each parameter are labeled at the top of the figure in degrees.
 \label{fig_mcmc}} 
\end{figure}

\begin{table}
 \centering
 \caption{ The three Gaussian fitting parameters for the integrated pulse profile of RRAT~J1541+4703. The errors result from least squares fitting of the Gaussian functions.  \label{table_guass}}
 {\begin{tabular}{ccccc}
  \hline
  \hline
   ith & $\rm I_i$ & $\rm x_i(^\circ)$ & $\rm w_i(^\circ)$ \\
  \hline
   1 & $0.46\pm0.01$ & $-9.2\pm0.1$ & $8.4\pm0.2$\\
   2 & $0.68\pm0.02$ & $4.8\pm0.4$ & $13.1\pm0.7$\\
   3 & $0.80\pm0.05$ & $14.2\pm0.2$ & $9.7\pm0.2$\\
  \hline
  \hline	
 \end{tabular}}
\end{table}

\begin{figure} 
 \centering
 \includegraphics[width=0.49\textwidth]{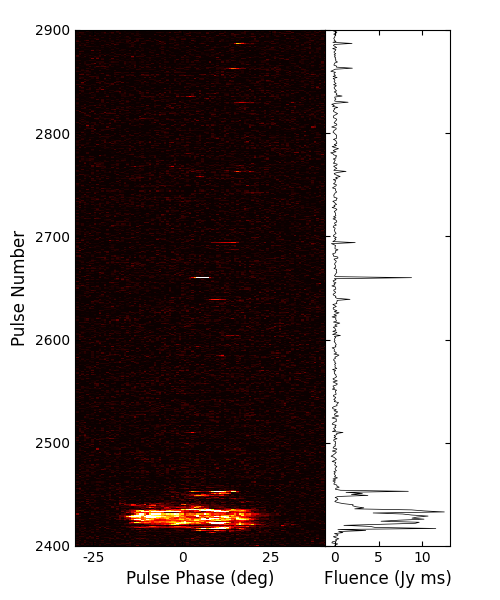}
 \caption{
 Single pulse stack of 500 consecutive pulses from RRAT~J1541+4703, exhibiting both RRAT and pulsar emission states. The right panel shows the fluence for each pulse, with distinct null pulses observed in both states.
 \label{fig_single_exm}} 
\end{figure}

\begin{figure*} 
 \centering
 \includegraphics[width=0.99\textwidth]{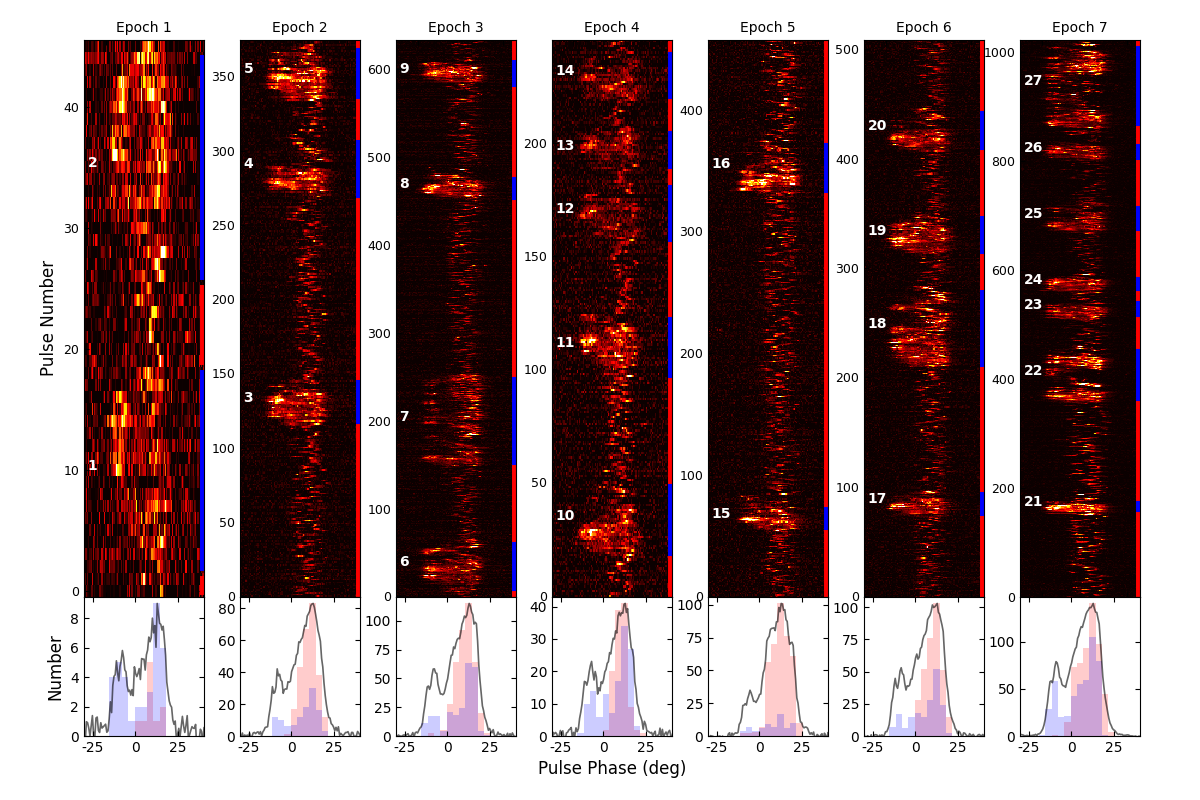}
 \caption{
 Top panel: Single pulse stack of burst pulses detected during seven observations of RRAT~J1541+4703. Bottom panel: Longitude distribution of single pulse peak phases for RRAT and NP states, shown by red and blue histograms, respectively. The black curve represents the integrated pulse profile. Note that all null pulses have been removed.
 \label{fig_burstsingle}} 
\end{figure*}


\subsection{Two emission states} \label{subsec_twostate}

Our observations reveal that the radio emission from RRAT~J1541+4703 exhibits both RRAT and NP states, detected in all seven observations. 
The most distinct difference between RRAT and NP states is that RRATs exhibit narrow, sparsely occurring bursts of short duration (typically one pulse periods), whereas NP states display broader pulses and denser bursts, often lasting tens to over a hundred pulse periods. Based on this characteristic, we identified their boundary using the following procedure. First, pulse sequences with durations exceeding five periods are preliminarily marked as NP states. Adjacent NP sequences separated by several null pulses were merged into a single event to establish preliminary NP boundaries. Next, the \citet{2018ApJ...866..152Shapiro-Albert} method was employed to perform multi-component Gaussian fitting (up to five components) on each burst pulse, estimating their $\rm W_{10}$. Finally, if a burst pulse occurs within 20 pulse periods of the NP state boundary, measure the $\rm W_{10}$ of the nearest burst pulse. If its $\rm W_{10}$ exceeds one-third of the integrated pulse profile width, classify it as an NP state. Simultaneously define a new boundary and continue this step until the NP state boundary is found.

Figure~\ref{fig_single_exm} shows a representative example, with the left panel displaying a time series of 500 consecutive single pulses and the right panel showing the corresponding fluence. 
The fluence for each pulse is calculated by integrating the pulse flux within the pulse window. 
As shown in Figure~\ref{fig_single_exm}, the emission is in RRAT state between 2500 and 2900 pulses, while it switches to the NP state around 2430 pulses. 
This phenomenon of coexistence has been observed in other pulsars \citep{2010MNRAS.402..855Burke, 2012ApJ...759L...3Esamdin, 2023ApJ...959...56SunSN}, and our findings contribute to expanding the sample of RRATs with this behavior.

The top panel of Figure~\ref{fig_burstsingle} shows the stack of single burst pulses detected across seven epochs, with all null pulses removed. The RRAT and NP states are clearly distinguishable, with boundaries marked by bold vertical red and blue lines on the right. Over the 7.4-hour observation, 27 bursts in NP state were detected (indicated by white digits on the left side of each figure), with transitions from RRAT to NP state occurring approximately every $\sim 0.3$ hour. 
The highest transition rate occurred during Epoch 7, with seven transitions within one hour. NP states were generally brief, with the longest lasting 170 pulse periods and the shortest just 21. In contrast, RRAT states lasted between 180 and over 8000 pulse periods, separating adjacent NP states. Additionally, burst pulses in the NP state were intermittently interrupted by random null pulses. The number of null and burst pulses detected in both states is listed in Table~\ref{table_obspara}. 
The results show that the null fraction (NF) for the RRAT state exceeds 95\%, with the highest reaching 99.8\%. In contrast, the NF for the NP state is much lower, with the highest value being 35.3\%.

Interestingly, the pulse emission window in the RRAT state is narrower than in the NP state and shifts toward the trailing portion of the integrated pulse profile. The bottom panel of Figure~\ref{fig_burstsingle} shows the longitudinal distribution of peak phases for single pulses in RRAT and NP states, represented by red and blue, respectively. The black curve represents the integrated pulse profile, normalized to the maximum value of the histogram. In the RRAT state, the peak of single pulses is concentrated at the tail of the profile, whereas in the NP state, the peaks are spread across the emission window, with most still occurring at the tail end.

The second and third rows of Figure~\ref{fig_mean_PA} show the polarized integrated pulse profile and corresponding PA for NP and RRAT states of RRAT~J1541+4703 at each epoch. Interestingly, the integrated pulse profile in the NP state resembles that of all burst pulses, with the first and second components being more prominent. In contrast, the RRAT state exhibits a narrower profile, with its peak occurring at a later phase. The integrated profile more closely resembles a Gaussian shape.
We estimated the widths of the two states. For the NP state, its pulse width is $\rm W_{50}=30.7^{\circ}\pm0.3^{\circ}$ and $\rm W_{10}=39.2^{\circ}\pm0.4^{\circ}$. For the RRAT state, its pulse width is $\rm W_{50}=14.4^{\circ}\pm1.5^{\circ}$ and $\rm W_{10}=23.3^{\circ}\pm2.1^{\circ}$. 
It is worth noting that the average PA of these two states differs, falling on RVM curves separated by $90^{\circ}$.
In the NP state, the PA follows the integrated profile, exhibiting $90^{\circ}$ jumps at phases $0^{\circ}$ and $20^{\circ}$, indicating the presence of two OPMs. In contrast, the RRAT state shows no $90^{\circ}$ jumps and has an average PA difference of $90^{\circ}$ compared to the NP state. While the single-pulse PA in the RRAT state suggests the presence of two OPMs, the average PA tends to favor one dominant mode. The linear polarization fractions for both states are nearly identical, measuring $16.8\% \pm 5.4\%$ for the NP state and $17.2\% \pm 6.2\%$ for the RRAT state.


\subsection{Single pulse feature} \label{subsec_single}

Figure~\ref{fig_seven_single} shows six examples of single pulse polarization profiles with different shapes. Each pulse profile differs notably from the integrated profile, with corresponding PA variations. The first two samples are from the RRAT state, while the remaining four are from the NP state. Single pulses in the RRAT state are typically narrow, whereas those in the NP state generally have broader widths, though narrow pulses are also present. In the RRAT state, single pulses typically have one or two peaks with relatively narrow widths. In contrast, NP state pulses exhibit more diversity, ranging from narrow profiles (e.g., pulse 7634 in Epoch 7) to double-peak, triple-peak, or even multi-peak structures, as seen in the last three single-pulse profiles in Figure~\ref{fig_seven_single}.

Figure~\ref{fig_scatter} shows the fluence versus equivalent width distribution for RRAT J1541+4703 during epochs 2 - 7, where the equivalent width is defined as the ratio of fluence to $\rm S_{peak}$. Epoch 1 is excluded due to its short observation time, which yielded too few samples for reliable statistics. 
The scatter plot reveals a clear positive correlation between pulse fluence and equivalent width across the detected burst pulses. 
Notably, the emission separates into two distinct clusters in this parameter space: one corresponding to RRAT state (red points) and the other to the NP state (blue points). 
The NP state predominantly occupy the region of higher fluence and broader widths, while the RRAT state pulses are concentrated at lower fluence and narrower widths. The marginal density distributions (top and right panels) further quantify this bimodality, showing peaked distributions for each state along both axes. 
This segregation suggests different underlying emission mechanisms or energetics for the two observed states.

\begin{figure*} 
 \centering
 \includegraphics[width=0.99\textwidth]{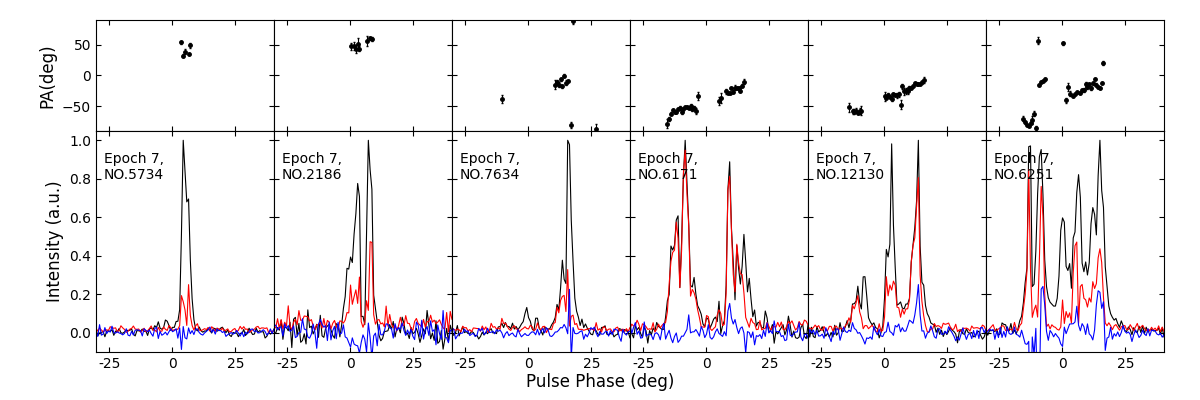}
 \caption{
 Polarization pulse profiles and PA of six single pulses from RRAT~J1541+4703. 
 \label{fig_seven_single}} 
\end{figure*}

\begin{figure*}
 \centering
 \includegraphics[width=0.99\textwidth]{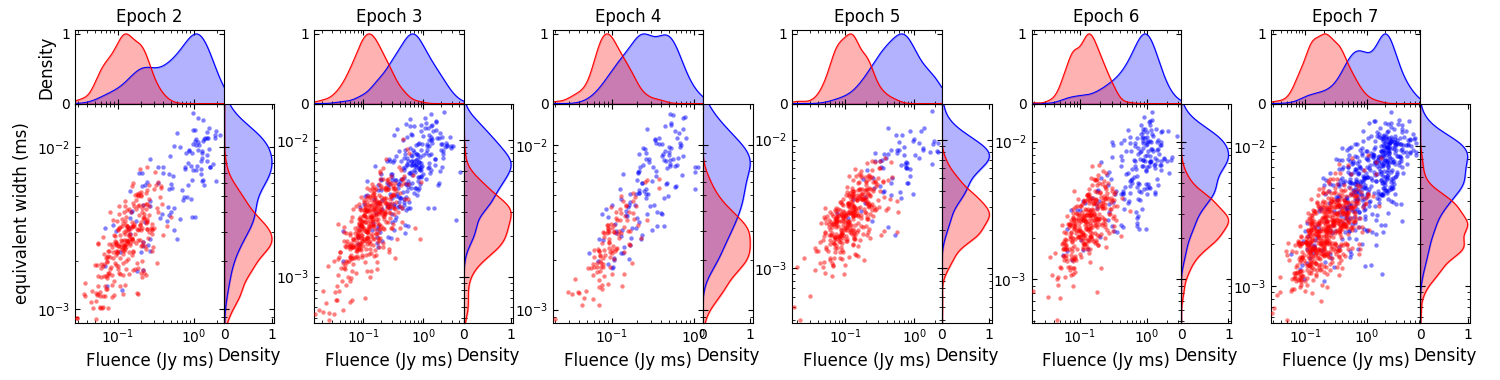}
 \caption{
 The fluence and equivalent width of RRAT~J1541+4703 during epochs 2 - 7. For each epoch, individual bursts are shown as a scatter plot, with the RRAT state in red and the NP state in blue. The top and right panels display the marginal density distributions for each parameter. 
 \label{fig_scatter}} 
\end{figure*}


\subsection{Waiting time Distribution} \label{subsec_pol}

The waiting time, or the duration between two consecutive pulses, provides an observational probe into the emission process of RRATs.
We measured the distribution of waiting times between consecutive burst pulses in the RRAT state across all epochs. Figure~\ref{fig_waiting_time} shows results for Epochs 2–7, with Epoch 1 omitted due to its short duration and insufficient sample. Except for Epoch 4, the waiting time distributions are generally consistent across epochs; deviations in Epoch 4 likely result from interference suppressing weak bursts. 
We assessed the goodness-of-fit for four probability distributions using the Kolmogorov–Smirnov (KS) test. For all epochs except Epoch 4, the KS p-values for the exponential distribution above the 0.05 confidence level, indicating no significant difference between the model and the observed data. This is consistent with the findings of some RRATs \citep{2023MNRAS.524.5132Dong, 2024MNRAS.527.4129ZhongWQ}. In contrast, some weaker bursts went undetected in Epoch 4, leading to deviations from the true distribution.
Furthermore, although the p-value derived from the Weibull distribution is below 0.05, the results indicate that the fit is acceptable. In contrast, Both the power‑law and gamma distributions produce p‑values below $10^{-10}$, indicating their inability to adequately describe the waiting time distribution of this RRAT.
We further compared the models using the AIC and BIC. The exponential distribution achieved the lowest AIC and BIC values compared to the Weibull distribution, indicating that it represents the optimal trade-off between goodness‑of‑fit and model complexity.
Taken together, both the KS p‑values and the information criteria consistently favor the exponential model. This result suggests that the waiting times of burst pulses are consistent with a Poisson process, implying no significant temporal correlation between successive pulses.
Furthermore, the waiting time distribution remains largely unchanged under reasonable threshold variations, confirming the robustness of our conclusion that the waiting times of burst pulses follow a Poisson process.

\begin{figure*} 
 \centering
 \includegraphics[width=0.99\textwidth]{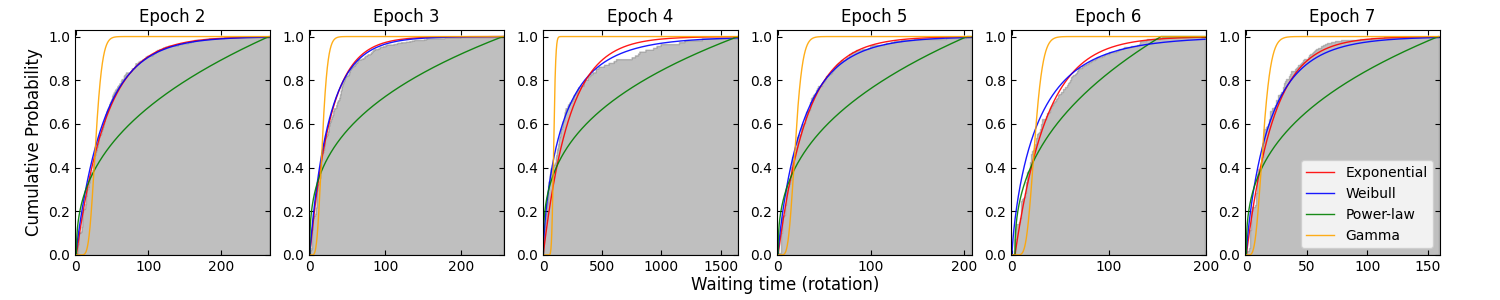}
 \caption{
 Cumulative probability distribution of single pulse wait times for RRAT~J1541+4703 from epoch 2 to epoch 7. The empirical data are compared to fits from four theoretical distributions: Weibull (red), gamma (blue), exponential (green), and power-law (yellow).
 \label{fig_waiting_time}} 
\end{figure*}


\subsection{Mode Changing and Subpulse Drifting} \label{subsec_modechange}

The shape of a pulsar's integrated profile remains stable at a given frequency \citep{1988MNRAS.234..477Lyne, 2017ApJ...845..156ZhaoRS}, exhibiting frequency dependent features \citep{1983ApJ...274..333Rankin, 2021ApJ...917..108XuX, 2016ApJ...816...76LuJG}. However, the emission features of some pulsars vary over time, causing the profile to switch between different modes, this phenomenon known as mode changing \citep{1970Natur.228.1297Backer}.
Based on the high sensitivity of FAST, we observed that NP state of RRAT~J1541+4703, we detected distinct mode changing behavior. Figure~\ref{fig_mode_hot} shows an example of a mode changing event, comprising a total of 73 consecutive single pulses. The emission switches between two modes, which we refer to as the normal and abnormal modes, distinguished by their occurrence frequency. The normal mode, more common, is denoted as N, while the abnormal mode is denoted as Ab in the figure.

The single pulse stacking example in Figure~\ref{fig_mode_hot} reveals the most distinct difference between the two modes: the intensity of the first component suddenly drops to near zero under abnormal modes. Based on this, to determine pulse profile, polarization, and timescale information for different modes, We use the average fluence within the phase range of the first component, $\rm S_{bin} = \sum_{i}^{N} S_i/N$, as a quantitative metric to distinguish between the two modes. 
Here, $\rm i$ denotes the phase, whose range is determined by 10\% of the first Gaussian component through the integration profile. $\rm N$ represents the number of phase bins.
The metric value for each single pulse is shown in the right panel of Figure~\ref{fig_mode_hot}. We use three times the root mean square of the off-pulse region as the threshold for distinguishing between the two modes, indicated by the green line in the figure. Pulses with metric values above this threshold are classified as normal mode, while those below are classified as abnormal mode.
Based on the above distinctions, we found that out of a total of 28 NP states, $\sim 44\%$ were in normal mode, while $\sim 39\%$ were in abnormal mode. The remainder are null pulses during the NP state.

Figure~\ref{fig_mode_profile} displays the polarization integrated profiles for the normal mode and abnormal mode in epoch 7.  The total intensity, linear polarization, and circular polarization for the normal and abnormal modes are shown by solid and dashed lines in black, red, and blue, respectively. The differences between the two modes are clearly evident. 
The top panel shows the PA swing for the normal mode (red error bars) and abnormal mode (blue error bars). In the normal mode, the first component is more prominent, even surpassing the third component, while in the abnormal mode, it is significantly weaker. Additionally, the linear and circular polarization, particularly at the pulse phase of the first component, differ between the two modes, and their PA values also show discrepancies.
Additionally, we observe that most NP states initially exhibit radiation in an abnormal mode, which then switches to a normal mode before reverting to the abnormal mode at the end. Multiple switching events between these modes may occur, potentially reflecting a sequence of changes within the pulsar's magnetosphere.

We note significant similarities between the abnormal mode and the RRAT state. Both exhibit narrower pulse widths and delayed peak phases compared to the normal mode. However, Figures~\ref{fig_mean_PA} and \ref{fig_mode_profile} show that their PA behaviors differ markedly: the abnormal mode retains frequent $90^{\circ}$ jumps similar to the normal mode, whereas the RRAT state exhibits a more stable PA swing. Moreover, in the integrated pulse profile, the first component of the abnormal mode is significantly weakened but not entirely absent, differing from the RRAT state. We suspect that the abnormal mode may represent an intermediate stage in the evolution from the normal mode to the RRAT state, though further observations are needed to confirm this hypothesis.

Previous research \citet{2023MNRAS.524.5132Dong} discovered that the RRAT~J1541+4703 exhibits subpulse drift behavior. Our observations reveal no significant subpulse drifting. Only a small fraction of subpulses show vague drifting behavior, with occasional phase movement. For example, pulses 6330–6335 in the main panel of Figure~\ref{fig_mode_profile} appear to shift toward earlier phases as subpulse phases progress. However, this behavior is rare and may be a random fluctuation. Thus, we suggest that RRAT~J1541+4703 does not exhibit subpulse drifting.

\begin{figure} 
 \centering
 \includegraphics[width=0.48\textwidth]{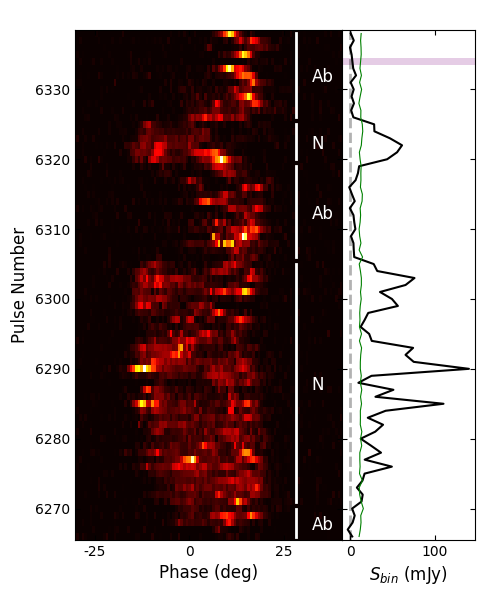}
 \caption{
 A sequence of 73 consecutive single pulses from RRAT~J1541+4703 in its NP state, demonstrating mode changing behavior. The right panel shows the $\rm S_{bin}$ values (black) and separation thresholds (green) for each pulse, with purple horizontal fills indicating null pulses.
 \label{fig_mode_hot}} 
\end{figure}

\begin{figure} 
 \centering
 \includegraphics[width=0.48\textwidth]{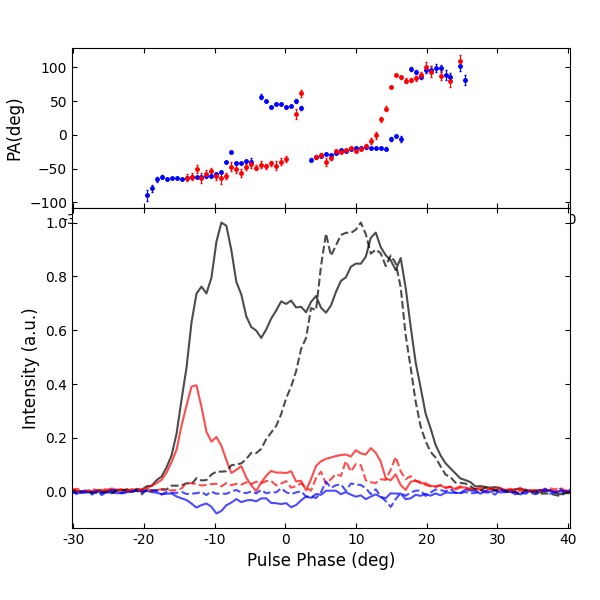}
 \caption{
 The average polarization properties of the normal (solid lines) and abnormal (dotted lines) modes for Epoch 7. The PA swings for the normal and abnormal modes are represented by blue and red error bars, respectively.
 \label{fig_mode_profile}}
\end{figure}


\section{Discussion and Conclusions} \label{sec_dis_con}

The classification of RRATs has been a key question. Timing analyses of some RRATs confirm that they are indeed periodic rotators \citep{2011MNRAS.415.3065Keane, 2015ApJ...809...67Karako-Argaman, 2023MNRAS.524.5132Dong}.
This indicates that RRAT most likely represents a member of the pulsar population \citep{2011MNRAS.415.3065Keane, 2010MNRAS.402..855Burke}.
Its sporadic emission features make it more similar to extreme nulling pulsars with long period \citep{2006Natur.439..817McLaughlin}, with NF values reaching as high as 95\% or more \citep{2020MNRAS.491..725Michilli, 2022ApJ...940L..21XieJT}.
Only a handful of RRATs exhibit switching between a sustained NP state and a sporadic RRAT state. Studying such rare objects is crucial for understanding the radio emission mechanisms of both pulsars and RRATs \citep{2006ApJ...645L.149Weltevrede, 2010MNRAS.402..855Burke, 2011MNRAS.411.1917Weltevrede}. Our observations of RRAT~J1541+4703 with FAST reveal it to be a new member of this special class, providing a key sample for this investigation.

We have, for the first time, directly observed RRAT~J1541+4703 switching between these two states. It spends ~98\% of the time in the RRAT state, punctuated by short-lived NP state episodes lasting 20–170 periods.
This RRAT may represent a connection between pulsars and RRATs, serving as an intermediate evolutionary stage from nulling pulsars to RRATs \citep{2010MNRAS.402..855Burke}. The discharge of sparks above the pulsar's polar cap causes high energy pairs to flow outward along magnetic field lines, generating radio emissions \citep{1975ApJ...196...51Ruderman}. In the RRAT state, sporadic bursts indicate the pair production is unstable in this state \citep{2007MNRAS.374.1103ZhangB}. In contrast, the NP state is significantly more stable, though this stability does not extend for very long.

Currently, most RRATs exhibit relatively low pulse rates, with some registering as few as several pulses per hour \citep{2011MNRAS.415.3065Keane, 2016ApJ...821...10Deneva, 2021ApJ...922...43Good}. Only a small fraction exceed 400 pulses per hour, such as, RRATs~J1439+76 and J1647$-$3607\citep{2010MNRAS.402..855Burke, 2015ApJ...809...67Karako-Argaman}. 
However, thanks to FAST's ultra-high sensitivity, pulse rates of certain RRATs have been found to be several times higher than previously estimated \citep[e.g. ][]{2022ApJ...940L..21XieJT, 2024MNRAS.528.1213DangSJ, 2024MNRAS.527.4129ZhongWQ}. 
Some RRATs have even been questioned under high-sensitivity observation as to whether they belong to RRAT group \citep{2025PASA...42..165YangT}. For RRAT~J1541+4703, previous studies reported a pulse rate of around $\sim 8$ pulses per hour \citep{2023MNRAS.524.5132Dong}. 
Our observations reveal a significantly higher rate, with a peak of 589 pulses per hour, making it the highest-pulse-rate RRAT discovered to date. If pulses in the NP state are included, the pulse rate will be higher. This suggests that pulse rate measurements are highly dependent on telescope sensitivity, and the burst rates of many RRATs should be re-evaluated.

Research indicates that the nulling phenomenon is an inevitable process in a pulsar's death journey, so nulling pulsars with high NF value tend to be in the death valley \citep{2000ApJ...531L.135ZhangB}. 
Compared to most nulling pulsars, RRATs exhibit higher NF values, placing them in this region, potentially approaching the bottom boundary of the Death Valley.
However, this is not always the case. Some RRATs are located at the upper boundary of Death Valley, or even farther \citep{2022JApA...43...75Abhishek}. RRAT~J1541+4703, for example, lies outside Death Valley, challenging current theories and necessitating a reexamination of the evolutionary process of pulsars.

For RRATs, if their bursts follow a Poisson process, the waiting times would be exponentially distributed, suggesting that RRAT radiation is a random process uniformly distributed over time. This behavior has been observed in some RRATs \citep[e.g. ][]{2023MNRAS.518.1418Hsu, 2024MNRAS.527.4129ZhongWQ, 2025ApJ...988...11DangSJ}.
Our results indicate that the waiting times for RRAT~J1541+4703 follow an exponential distribution. In contrast, repeating fast radio bursts conform to a Weibull distribution \citep{2018MNRAS.475.5109Oppermann}. 
For NP states, our results show an average occurrence every 0.3 hours. However, within the limited observation time, we cannot determine a reliable distribution of waiting times between sequences. Extending the observation duration is necessary to ascertain this.

The RRAT~J1541+4703 exhibits higher and broader for single pulse fluence distribution in NP state, with a wider and more complex pulse profile. Most single pulses in RRAT mode are narrow and single peaked or double peaked, with fluence fluctuating several times around the distribution peak, resembling some RRATs \citep{2024ApJ...976L..26ZhangSB}.
However, this is not true for all RRATs; some even span three orders of magnitude \citep{2025MNRAS.539.1352TangZF}.
Within the framework of the Partially Screened Gap (PSG) model \citep{2000ApJ...541..351Gil, 2006ApJ...650.1048Gil}, the polar cap region of pulsars exhibits two distinct modes: the PSG-on mode and the PSG-off mode \citep{2015MNRAS.447.2295Szary}. These may correspond to the two states demonstrated by RRATs.
In this case, the PSG-off mode exhibits higher fluence due to its greater gap, corresponding to radiation from the NP state, while the PSG-on mode produces lower fluence corresponding to the RRAT state. 

For the integrated polarized pulse profile, the NP state exhibits three components with a $90^{\circ}$ jump in the average PA. In contrast, the RRAT state possesses only one major component with a stable average PA. The pulse phase shift between the two states at the contour center is $\sim 6.5^{\circ} \pm 1.3^{\circ}$. This exhibits features similar to those of RRAT~J1752+2359 \citep{2021RAA....21..240SunSN}.
Research indicates that the emission beams of pulsars consist of a central component and two nested cone components \citep{1983ApJ...274..333Rankin}, each with differing emission heights \citep{2001A&A...377..964QiaoGJ, 2022ApJ...926...73ZhiQJ}. The delay effect causes the beam center to shift backward \citep{1992ApJ...385..282Phillips}, resulting in an asymmetric profile and potentially leading to jumps in the average PA \citep{1997A&A...323..395XuRX}.
This may also suggest the combined effect of two plasma propagation modes within the magnetosphere \citep{1986ApJ...302..120Arons}.

Mode changing is another intriguing characteristic of RRAT~J1541+4703, observed during NP states. The most notable difference is a weaker first component in the abnormal mode.
Additionally, the PA in the abnormal mode shows slight deviations from that of the normal mode, accompanied by a significant reduction in the polarization fraction. This suggests changes in the magnetospheric physical processes during the NP state. This phenomenon is not unique to J1541+4703. 
Similar mode changing behavior has been observed in other pulsars, such as PSRs~J1326$-$6700 and B2020+28. In J1326$-$6700, the abnormal mode also exhibits reduced intensity in specific emission components \citep{2020ApJ...904...72WenZG}. In B2020+28, mode switching primarily manifests as a change in the relative intensity between the leading and trailing components \citep{2016Ap&SS.361..261WenZG}.
Some studies suggest that mode changing may be triggered by fluctuations in the surface temperature of pulsars \citep{1997ApJ...478..313ZhangB}, magnetic field instabilities \citep{2003A&A...412L..33Geppert}, variations in currents within the magnetosphere \citep{2010Sci...329..408Lyne}, or differences in emission heights \citep{2020ApJ...904...72WenZG}. However, no definitive conclusion has been reached, and further investigation is required.
It is noteworthy that the abnormal mode in the NP state exhibits a pulse profile similar to that of the RRAT state. The primary difference is that the leading component in the abnormal mode does not completely disappear, and there are significant differences in polarization, indicating two distinct magnetospheric dynamics. We speculate that the abnormal mode may represent an intermediate evolutionary stage between the normal mode and the RRAT state, which requires verification through future observations.


The last but not the least, it is a real big issue why the sparking points distribute non-symmetrically to the meridian co-plane of the spin, the magnetic axis and the line of sight. The offset of the emission peaks from the RRAT state implies that the pair production site on the pulsar’s polar cap surface is more likely to be located toward the tail. This may indicate that the surface of the pulsar is not flat, with small “zits” along its tail \citep{2026RAA....26c5014XuZH}, if pulsars are strangeon stars \citep{2003ApJ...596L..59XuRX, 2023AN....34430008XuRX}. 
The priority discharge of sparks is related to the height of “zits”, where the parallel electric field around the peaks is enhanced, thereby achieving more efficient electron acceleration. This surface condition is also considered in explaining radio emissions, including mode changing and diffuse subpulse drifting \citep{2019SCPMA..6259505LuJG, 2024ApJ...973...56CaoSS, 2024ApJ...963...65WangZL}. If this is the case, the emission properties of RRAT J1541+4703 provide stronger support for the case where the pulsar is a strangeon stars.



 
\begin{acknowledgments}

This work made use of the data from FAST (Five-hundred-meter Aperture Spherical radio Telescope)(https://cstr.cn/31116.02.FAST). FAST is a Chinese national mega-science facility, operated by National Astronomical Observatories, Chinese Academy of Sciences.
This work is supported by the National Natural Science Foundation of China (Nos. 12273008), the National SKA Program of China (Nos.2022SKA0130100, 2022SKA0130104), the Natural Science and Technology Foundation of Guizhou Province (Nos. [2023]024, ZK[2022]304, KY[2022]132), the Major Science and Technology Program of Xinjiang Uygur Autonomous Region (No.2022A03013-4).

\end{acknowledgments}

\bibliography{references}{}
\bibliographystyle{aasjournalv7}

\end{document}